\documentclass[aps,prc,preprint]{revtex4}
\usepackage{amsfonts}
\usepackage{amssymb}
\usepackage{graphics}
\usepackage{graphicx}
\usepackage[usenames,dvipsnames]{color}
\newcommand{\la}{\langle}
\newcommand{\ra}{\rangle}
\newcommand{\ls}{\langle\sigma\rangle}
\newcommand{\lw}{\langle\omega\rangle}
\newcommand{\lr}{\langle\rho\rangle}

\begin{document}

\title{Neutron matter under strong magnetic fields: a
comparison of models}

\author{R. Aguirre$^{1,2}$, E. Bauer$^{2,3}$ and I. Vida\~na$^4$}
\email{aguirre@fisica.unlp.edu.ar, bauer@fisica.unlp.edu.ar, ividana@fis.uc.pt}
\affiliation{$^1$Departamento de Fisica, Facultad de Ciencias Exactas, Universidad Nacional de La Plata, Argentina}
\affiliation{$^2$ IFLP, CCT-La Plata, CONICET. Argentina}
\affiliation{$^3$Facultad de Ciencias Astron\'omicas y Geof\'{\i}sicas, Universidad Nacional de La Plata, Argentina}
\affiliation{$^4$Centro de F\'{\i}sica Computacional. Department of Physics. University of Coimbra, PT-3004-516 Coimbra, Portugal}

%%%%%%%%%%%%%%%%%%%%%%%%%%%%%%%%%%%%%%%%%%%%%%%%%%%%%%%%%%%%

\begin{abstract}
The equation of state of neutron matter is affected by the
presence of a magnetic field due to the intrinsic magnetic moment
of the neutron. Here we study the equilibrium configuration of
this system for a wide range of densities,
temperatures and
magnetic fields. Special attention is paid to the behavior of the
isothermal compressibility and the magnetic susceptibility.
Our calculation is performed using both microscopic and
phenomenological approaches of the neutron matter equation of
state, namely the Brueckner--Hartree--Fock (BHF) approach using
the Argonne V18 nucleon-nucleon potential supplemented with the
Urbana IX three-nucleon force, the effective Skyrme model in a
Hartree--Fock description, and the Quantum Hadrodynamic
formulation with a mean field approximation. All these approaches
predict a change from completely spin polarized to partially
polarized matter that leads to a continuous equation of state. The
compressibility and the magnetic susceptibility show
characteristic behaviors, which reflect that fact. Thermal effects
tend to smear out the sharpness found for these quantities at T=0.
In most cases a thermal increase of $\Delta T=10$ MeV is enough to
hide the signals of the change of polarization. The set of
densities and magnetic field intensities for which the system
changes it spin polarization is different for each model. However,
we found that under the conditions examined in this
work there is an overall agreement between the three theoretical
descriptions.
\end{abstract}


\maketitle

%%%%%%%%%%%%%%%%%%%%%%%%%%%%%%%%%%%%%%%%%%%%%%%%%%%%%%%%%%%%%%
\section{Introduction}

The effects of magnetic fields on dense matter have been a subject
of interest for a long time (see {\it e.g.,} Ref.\ \cite{LAI} and
references therein), particularly in relation to astrophysical
issues. The equation of state for magnetized matter is important
for the neutron star structure~\cite{LATTIMER} and for the cooling
of magnetized stars~\cite{SHIBANOV, YAKOVLEV, BEZCHASTNOV}.
Moreover, since neutrinos have a fundamental role in cooling
processes, their emission and transport properties in the presence
of magnetic fields have also been studied in detail
\cite{YAKOVLEV, BEZCHASTNOV}. A wide range of observational data
of periodic or irregular radiation from localized sources has been
related to the presence of very intense magnetic fields in compact
stellar objects. These manifestations have been associated with
pulsars, soft gamma ray repeaters and anomalous X-ray pulsars,
according to the energy released and the periodicity of the
episodes. Thus, they have been associated with different stages of
the evolution of neutron stars. The intensity of the magnetic
fields could reach values, {\it e.g.,} in the case of magnetars,
up to $10^{14}-10^{15}$ G in the star surface and grows by several
orders of magnitude in its dense interior. The origin of such
unusually large intensities is still uncertain. A possible
explanation invokes a spontaneous phase transition to a
ferromagnetic state at densities corresponding to theoretically
stable neutron stars and therefore a ferromagnetic core in the
liquid interior of such compact objects. Such
possibility has been considered since long ago by many authors
within different theoretical approaches (see {\it e.g.,} Refs.\
\cite{BROWNELL,RICE,CLARK,CLARK2,SILVERSTEIN,OST,PEAR,PANDA,BACK,HAENSEL,JACK,KUT,MARCOS,
VIDAURRE,RIOS,GOGNY,FANTONI,VIDANA,VIDANAb,VIDANAc,SAMMARRUCA1,SAMMARRUCA2,BIGDELI}),
but results were contradictory. Whereas some calculations based
on Skyrme \cite{VIDAURRE,RIOS} or Gogny \cite{GOGNY} interactions
predicted such a transition to occur at densities in the range
$(1-4)n_0$ ($n_0=0.16$ fm$^{-3}$), other calculations that used
modern two- and three-body realistic interactions, like Monte
Carlo \cite{FANTONI}, Brueckner--Hartree--Fock (BHF)
\cite{VIDANA,VIDANAb,VIDANAc}, Dirac--Brueckner--Hartree--Fock
\cite{SAMMARRUCA1,SAMMARRUCA2}, or lowest-order constraint
variational \cite{BIGDELI}, excluded such a transition.
Nowadays, there is a general consensus that the ferromagnetic instability
predicted by the Skyrme forces at high densities is in fact a
pathology of such forces. Modifications of the standard Skyrme
interaction have been recently proposed \cite{SAGAWA,CHAMEL} in
order to remove the instability.

Recently it was pointed out \cite{KHARZEEV,MO,SKOKOV} that matter
created in heavy ion collisions could be subject to very strong
magnetic fields, with distinguishable consequences
on particle production. In Ref. \cite{KHARZEEV} a magnetic
field is predicted in noncentral heavy ion collisions such that
$e\, B \sim 10^2$ MeV$^2$, which would be the cause of a preferential
emission of charged particles along the direction of the magnetic
field. Improvements in the description of the mass distribution of
the colliding ions \cite{MO} does not modify essentially the
magnitude of the fields produced.  On the other hand, the
numerical simulations performed by \cite{SKOKOV} predicts values
$e\, B \sim m_\pi^2$ MeV$^2$, which are much larger
than those in Ref.\ \cite{KHARZEEV}.

%%%%%%%%%%%%%%%%%%%%%%%%%%%%%%%%%%%%%%%%%%%%%%%%

Several models have been used to describe the effects of magnetic
fields in a dense nuclear environment and particularly on the
properties and the structure of neutron stars \cite{CHAKRABARTY,
BRODERICK, YUE2,SINHA, RABHI2,RABHI0b,RYU,RYU2,RABHI3,
ANG,DONG,DIENER,DONG2,PGARCIA, ISAEV2,UNLP,BORDBAR,BIGDELI2}.
Among them, covariant field theoretical models have been
extensively used to study the role of the magnetic field in
hyperonic matter \cite{YUE2,SINHA}, instabilities at subsaturation
densities \cite{RABHI2,RABHI0b,RYU}, magnetization of stellar
matter \cite{DONG}, saturation properties of symmetric matter
\cite{DIENER}, and the symmetry energy \cite{DONG2}.
Non-relativistic models have also been used in this context
\cite{PGARCIA, ISAEV2,UNLP} and microscopic models based on
realistic nucleon forces, such as the recent lowest-order
variational calculations of Refs.\ \cite{BORDBAR} and
\cite{BIGDELI2}.

Due to its important applications, as well as to its intrinsic
theoretical interest, a comparison of predictions is in order.
Therefore, we have selected three models, of very different
origins,  to study the properties of infinite homogeneous neutron
matter in the presence of strong magnetic fields. They are the
Brueckner--Hartree--Fock (BHF) approach using the Argonne V18
nucleon-nucleon potential supplemented with the Urbana IX
three-nucleon force, the covariant formulation known as Quantum
Hadro-dynamics (QHD), and the non-relativistic Skyrme effective
potential. It would be interesting to include a comparison
with other microscopic calculations, as for instance with the auxiliary field
diffusion Monte Carlo (AFDMC) method \cite{FANTONI,AFDMC}. As an
interesting precedent, it must be mentioned that a comparison
between the BHF and AFDMC approaches was already done in
Ref.\ \cite{VIDANA}. The results for the magnetic
susceptibility of spin polarized neutron matter at zero
temperature in the absence of a magnetic field, give a remarkable
agreement between the two methods.

In the present work, we focus on the polarization of neutron matter
by analyzing its dependence with density, magnetic field and
temperature. In order to understand this behavior, we also consider
the energy of the system and its pressure. In addition, we pay
special attention to some thermodynamical coefficients: the
isothermal compressibility and the magnetic susceptibility. We
consider a range of densities where nucleons are the main degrees of
freedom of hadronic matter, and temperatures and field intensities up
to $T=10$ MeV  and $B=10^{19}$ G, respectively.

This article is organized as follows. A brief review of the  properties of
spin polarized neutron matter and of the models and approximations used is
presented in the next section, and the results are shown and discussed in Section
\ref{Results}. A final summary and the main conclusions are given in Section
\ref{Summary}.

%%%%%%%%%%%%%%%%%%%%%%%%%%%%%%%%%%%%%%%%%%%%%%%%%%%%%%%%%%%%

\section{Neutron matter in an external magnetic field}
\label{SecII}

Spin polarized neutron matter is an infinite nuclear system made of
two different fermionic components: neutrons with spin up and
neutrons with spin down, having number densities $n_\uparrow$ and
$n_\downarrow$, respectively. Hence, the total number density is
given by
\begin{equation}
n=n_\uparrow+n_\downarrow\ .
\label{totd}
\end{equation}
The degree of spin polarization of the system can be expressed by means of the
spin asymmetry density, defined as
\begin{equation}
W=n_\uparrow-n_\downarrow \ .
\label{sad}
\end{equation}
Note that the value $W=0$ corresponds to nonpolarized
neutron matter, whereas $W=n$ or $W=-n$
means, respectively, that the system is in a completely polarized state with all the spins up
(CPS-U) or down (CPS-D), {\it i.e.}, all the spins are
aligned along the same direction. Partially polarized states (PPS)
correspond to  values of $W$ between $-n$ and $n$.

In the following we present the main features of the three
approaches used to describe the properties of spin-polarized neutron
matter in the presence of an external magnetic field B. We evaluate
first, for the three approaches, the Helmhotz free-energy density of
the system {${\mathcal F}={\mathcal E}-T{\mathcal S} $}, where the
energy density ${\mathcal E}$ includes the term describing the
interaction of matter with the external field, and the entropy
density ${\mathcal S}$ is evaluated in the quasi-particle
approximation. Then, we determine from ${\mathcal F}$ other
macroscopic properties of the system such as the pressure $P$, the
magnetization of the system per unit volume
\begin{equation}
{\cal M}=\left(\frac{\partial P}{\partial B}\right)_{\mu,T,\Omega} \ ,
\end{equation}
the isothermal compressibility
\begin{equation}
K=-\frac{1}{\Omega}\left(\frac{\partial \Omega}{\partial P}
\right)_{N,T,B}
\end{equation}
where $\Omega$ is the volume of the system, and the magnetic susceptibility
\begin{equation}
\chi = \left(\frac{\partial {\cal M}}{\partial B}\right)_{N,T,\Omega}
\end{equation}
which characterizes the response of the system to the external field
and gives a measure of the energy required to produce a net spin
alignment in the direction of the field.

From here on we use units such that $\hbar=1$,
$c=1$, and $k_B=1$.
%%%%%%%%%%%%%%%%%%%%%%%%%%%%%%%%%%%%%%%%%%%%%%%%%%%
\subsection{The BHF approach}

The extension of the BHF approach for neutron matter in the
presence of a magnetic field and finite temperature starts with
the construction of the neutron-neutron $G$-matrix. It describes,
in an effective way, the interaction between two neutrons for any
spin combination.
This is formally obtained by solving the well known
Bethe--Goldstone equation
\begin{eqnarray}
&&\la{\vec k_3}\sigma_3, {\vec k_4}\sigma_4|G(\omega)|{\vec k_1}\sigma_1, {\vec k_2}\sigma_2\ra=
\la{\vec k_3}\sigma_3, {\vec k_4}\sigma_4|V|{\vec k_1}\sigma_1, {\vec k_2}\sigma_2\ra \nonumber \\
&&+\frac{1}{\Omega}\sum_{\sigma_i{\vec k_i},\sigma_j{\vec k_j}}
\la{\vec k_3}\sigma_3, {\vec k_4}\sigma_4|V|{\vec k_i}\sigma_i, {\vec k_j}\sigma_j\ra
\frac{Q_{\sigma_i\sigma_j}({\vec k_i},{\vec k_j}) }{\omega-\epsilon_{\sigma_i}-\epsilon_{\sigma_j}+i\eta}
\la{\vec k_i}\sigma_i, {\vec k_j}\sigma_j|G(\omega)|{\vec k_1}\sigma_1, {\vec k_2}\sigma_2\ra
\label{bgeq}
\end{eqnarray}
where $\sigma=\uparrow, \downarrow$ indicates the spin projection
of each neutron in the initial, intermediate and final states, $V$
is the bare nucleon-nucleon interaction,
$Q_{\sigma_i\sigma_j}({\vec k_i},{\vec
k_j})=(1-\nu_{\sigma_i}({\vec k_i}) )(1-\nu_{\sigma_j}({\vec k_j}))$,
where $\nu_{\sigma}({\vec k})$ is the occupation number defined in Eq.\ (\ref{ocn}),
is the Pauli operator
which allows only intermediate states compatible with the Pauli
principle,
and $\omega$ is the so-called starting energy defined
as the sum of the non-relativistic single-particles energies
$\epsilon_{\uparrow(\downarrow)}$ of the interacting neutrons.
Note that Eq.\ (\ref{bgeq}) is a coupled channel equation.

The single-particle energy of a neutron with momentum ${\vec k}$ and spin projection $\sigma$ in the presence of
an external magnetic field $B$ is given by
\begin{equation}
\epsilon_{\sigma}({\vec k})=\frac{%\hbar^2
k^2}{2m}+\mbox{Re}[U_{\sigma}({\vec k})]\mp \kappa B \ ,
\label{spe}
\end{equation}
where the real part of the single-particle potential
$U_{\sigma}({\vec k})$  represents the average potential ``felt'' by
a neutron in the nuclear medium. The minus (plus) sign in the last
term corresponds to neutrons with spin up (down), and
$\kappa=-1.913\mu_N$ is the anomalous magnetic moment of the neutron
with $\mu_N$ the nuclear magneton. In the BHF approximation
$U_{\sigma}({\vec k})$ is given by
\begin{equation}
U_{\sigma}({\vec k})=\frac{1}{\Omega}\sum_{\sigma',{\vec k'}}\nu_{\sigma'}({\vec k'})
\langle {\vec k}\sigma, {\vec k'}\sigma' |G(\omega=\epsilon_{\sigma}(\vec k)+\epsilon_{\sigma'}(\vec k'))|{\vec k}\sigma, {\vec k'}\sigma'\rangle_{A}
\label{spp}
\end{equation}
where the occupation number of a neutron with spin projection
$\sigma$  at zero temperature is
\begin{equation}
\nu_{\sigma}({\vec k})=
\left\{
\begin{array}{cc}
1, & \mbox{if} \,\, |{\vec k}| \leq k_{F_\sigma} \\
0, & \mbox{otherwise}
\end{array}
\right .
\label{ocn}
\end{equation}
with $k_{F_\sigma}=(6\pi^2n_\sigma)^{1/3}$ the corresponding Fermi
momentum, and the matrix elements are properly antisymmetrized.  We
note here that the so-called continuous prescription \cite{JEKEUNE}
has been adopted for the single-particle potential when solving the
Bethe--Goldstone equation. It has been shown by Song {\it et al.,}
\cite{SONG} that the contribution to the energy from {\it
three-hole-line} diagrams (which account for the effect of
three-body correlations) is
minimized when this prescription is adopted.
This presumably enhances the convergence of the hole-line expansion
of which the BHF approximation represents the lowest order. We
also note that the present BHF calculation has been carried out
using the Argonne V18 potential \cite{WIRINGA} supplemented with
the Urbana IX three-nucleon force \cite{PUDLINER}, which for the
use in the BHF approach is reduced first to an effective
two-nucleon density-dependent force by averaging over the
coordinates of the third nucleon \cite{TBF}.

The total energy per unit
volume is easily obtained once  a self-consistent solution of
Eqs.\ (\ref{bgeq})-(\ref{spp}) is achieved
\begin{equation}
{\mathcal E}=\frac{1}{\Omega}\sum_{\sigma{\vec k}}\nu_{\sigma}({\vec k})
\left(\frac{%\hbar^2
k^2}{2m}+\frac{1}{2}\mbox{Re}[U_{\sigma}(\vec k)]\right)
-\kappa W B \ ,
\label{bhfed}
\end{equation}
where $W$ is the spin asymmetry density defined in Eq.\ (\ref{sad}).

For further purposes, it is convenient to introduce the effective mass $m_{\sigma}^*(k)$
defined as,
\begin{equation}
\frac{m_{\sigma}^*(k)}{m}=\frac{k}{m}\left( \frac{d \epsilon_{\sigma}(k)}{dk} \right)^{-1} \ ,
\label{efm}
\end{equation}
where $m$ is the bare neutron mass.

At the BHF level, finite temperature effects can be introduced in
a very good approximation just by replacing the zero temperature
limit of the occupation number, Eq (\ref{ocn}), by its full
expression
\begin{equation}
f_{\sigma}({\vec k}, T)=\frac{1}{1+\exp[(\epsilon_{\sigma}({\vec k}, T)-\mu_{\sigma}(T))/T]} \ ,
\label{fdd}
\end{equation}
into the formulae shown in Eqs. (\ref{bgeq}), (\ref{spp}), and
(\ref{bhfed}). Here $\mu_{\sigma}(T)$ is the chemical potential of
a neutron  with spin projection $\sigma$.

These approximations, valid in the range of densities and
temperatures considered here, correspond to the ``naive'' finite
temperature Brueckner--Bethe--Goldstone expansion discussed by
Baldo and Ferreira in Ref.\ \cite{BF}. The interested reader is
referred to this work and references therein for a formal and
general discussion on the nuclear many-body problem at finite
temperature.

In this case, however, the self-consistent procedure implies that,
together  with the Bethe-Goldstone equation and the
single-particle potential, the chemical potentials of neutrons
with spin up and down must be extracted at each step of the
iterative process from the normalization condition
\begin{equation}
n_{\sigma}=\frac{1}{\Omega}\sum_{\vec k}f_{\sigma}({\vec k}, T) \ .
\label{nc}
\end{equation}
This is an implicit equation which can be solved numerically. Note
that the $G$-matrix obtained from the Bethe--Goldstone equation
(\ref{bgeq}) and also the single-particle potentials depend
implicitly on the chemical potentials.

Once a self-consistent solution is achieved,  the
entropy per unit volume is calculated in the quasi-particle
approximation
\begin{equation}
{\mathcal S}=-\frac{1}{\Omega}\sum_{\sigma{\vec k}}
\left(f_{\sigma}({\vec k}, T)\mbox{ln}(f_{\sigma}({\vec k}, T))
+(1-f_{\sigma}({\vec k}, T))\mbox{ln}(1-f_{\sigma}({\vec k},
T))\right) \ , \label{entrop}
\end{equation}
which together with the energy density ${\mathcal E}$ are used to evaluate
the Helmhotz free energy density ${\mathcal F}$.

Finally, for fixed values of the total density $n$, the
temperature $T$ and the external field $B$, the physical state is
simply obtained by minimizing ${\mathcal F}$ with respect to the
spin asymmetry density $W$. We note that this minimization implies
that in the physical state $\mu_\uparrow=\mu_\downarrow$, {\it i.e.,}
there is only one chemical potential  which is associated to the
conservation of the total baryonic number.

%%%%%%%%%%%%%%%%%%%%%%%%%%%%%%%%%%%%%%%%%%%%%%%%%%%%%%%%%%%%%%%%%%%%
\subsection{The QHD model}

QHD is a covariant formulation of field theory, where the nuclear
interaction is mediated by the exchange of the
following mesons: the scalar isoscalar $\sigma$-meson, the vector
isoscalar $\omega$-meson, and the vector isovector $\rho$-meson. We
adopt here the FSU-Gold model \cite{FSUGOLD}, where a
meson self-interaction is added to the ordinary nucleon-meson
vertices. The lagrangian density reads
\begin{eqnarray}
{\mathcal L}&=& \bar{\psi} \left(i \not \! \partial -m +g_\sigma
\sigma- g_\omega \not \! \omega - \frac{g_\rho}{2} \mathbf{\tau}
\cdot \not \! \mathbf{\rho} + \kappa \sigma_{\mu \nu} F^{\mu
\nu}\right) \psi + \frac{1}{2} (\partial^\mu \sigma
\partial_\mu \sigma - m_\sigma^2 \sigma^2)\nonumber\\&&-\frac{g_2}{3}\, \sigma^3-\frac{g_3}{4} \,\sigma^4
 -\frac{1}{4}
 W^{\mu \nu} W_{\mu \nu} + \frac{1}{2} m_w^2
\omega^2 +\frac{C}{4} \, \omega^4-\frac{1}{4}
 R^{\mu \nu}\cdot R_{\mu \nu} + \frac{1}{2} m_r^2
\rho^2 + D \rho^2 \; \omega^2\nonumber \nonumber
\end{eqnarray}

\noindent where  we have used $\omega^2=\omega^\mu
\omega_\mu,\;\rho^2=\mathbf{\rho}^\mu \cdot \mathbf{\rho}_\mu,\;
W_{\mu \nu}=\partial_\mu \omega_\nu-\partial_\nu \omega_\mu, \;
R_{\mu \nu}=\partial_\mu \rho_\nu-\partial_\nu \rho_\mu$,
$\sigma_{\mu \nu}=i[\gamma_\mu,\gamma_\nu]/2$,
and $F^{\mu \nu}$ is the electromagnetic
tensor. Furthermore $g_\sigma, g_\omega, g_\rho, g_2, g_3, C$,
and $D$ are the coupling constants.
We note that in the above expression  the index ``$\sigma$'' should not be confused with
the spin projection of the neutron.

In the mean field approximation the meson fields $\sigma, \omega, \rho$ are
replaced by their corresponding in-medium expectation values
$\ls, \lw$ and $\lr$ which obey the following equations of motion
\begin{eqnarray}
\left(i \not \! \partial -m +g_\sigma \ls - g_\omega\gamma_0  \lw +
\frac{1}{2}
g_\rho
\gamma_0  \lr  + \kappa \sigma_{\mu \nu} F^{\mu \nu}\right)\psi&=&0, \label{NUCLEONEQ} \\
m_\sigma^2 \ls + g_2 \ls^2 + g_3 \ls^3&=&  g_\sigma n_s
,\label{SIGMA} \\
 m_{\omega}^2 \lw+ C \lw^3 + 2 D \lr^2 \lw & =& g_\omega n, \label{OMEGA} \\%\nonumber\\
m_{\rho}^2 \lr+ 2 D \lw^2 \lr &=&- g_\rho  n \label{RHO}
%\nonumber
\end{eqnarray}
where

\begin{equation}
n=\la\bar{\psi} \gamma_0 \psi\ra = \frac{1}{\Omega}\sum_{\sigma\vec k}f_{\sigma}({\vec k}, T)
\end{equation}
is the total number density and
\begin{equation}
n_s=\la\bar{\psi} \psi\ra=\frac{1}{\Omega}\sum_{\sigma\vec k}\frac{m^*}{E_{\sigma}({\vec k})}f_{\sigma}({\vec k},T)
\end{equation}
is the scalar density, $m^*=m-g_{\sigma}\ls$ is the neutron
effective mass and
\begin{equation}
E_{\sigma}({\vec k})=\sqrt{k^2_{//}
+\left(\sqrt{m^{*\,2}+k^2_\bot} \mp\kappa B\right)^2 }
\end{equation}
the relativistic energy. Here one must distinguish the momentum
components parallel ($k_{//}$) and perpendicular ($k_\bot$) to the
magnetic field. As in Eq.\ (\ref{spe}) the minus (plus) sign in
the above expression is for neutrons with spin up (down). The
single-particle energy is given in this model by
\begin{equation}
\epsilon_{\sigma}({\vec k})=E_{\sigma}({\vec k})+g_\omega\lw-\frac{1}{2}g_\rho\lr \ ,
\end{equation}
which corresponds to one of the eigenvalues of Eq.\ (\ref{NUCLEONEQ}).

The energy per unit volume can be evaluated as the component
$T^{00}$ of the energy-momentum tensor

\begin{eqnarray}
{\mathcal E}&=&\frac{1}{\Omega}\sum_{\sigma\vec k}
f_{\sigma}({\vec k}, T)  E_{\sigma}({\vec k}) +
\frac{1}{2}\left((m_{\sigma} \ls)^2+(m_{\omega} \lw)^2+(m_{\rho}
\lr)^2
\right) \nonumber \\
&+&\frac{1}{3} g_2 \ls^3+\frac{1}{4} g_3  \ls^4+\frac{3}{4} C
\lw^4+3 D   (\lw \lr)^2 \label{QHD_E}
\end{eqnarray}

The magnetization has a simple expression
\begin{equation}
{\cal M}=\frac{\kappa}{\Omega}\sum_{\sigma\vec k} s
f_{\sigma}({\vec k}, T) \frac{\sqrt{m^{*2}+\hbar^2k_\bot^2} - s \kappa  B}{ E_{\sigma}({\vec k})}
\end{equation}
where $s=1(-1)$ for $\sigma=\uparrow(\downarrow)$.

%%%%%%%%%%%%%%%%%%%%%%%%%%%%%%%%%%%%%%%%%%%%%%%%%%%%
\subsection{The Skyrme model}

The Skyrme model is an effective formulation of the nuclear
interaction \cite{SKYRME}. It consists of a two-body
contact potential plus some terms having an explicit density
dependence which, in an effective way, represent the
effect  of the three- and multi-body forces. Using this
interaction in the Hartree--Fock approximation, one builds up an
energy density functional. The associated single--particle
spectrum can be expressed in such a way that the interaction
contributes partly to the definition of an effective mass and
partly to a remaining potential energy. In the present work we
adopt the SLy4 parametrization from Ref.\ \cite{DOUCHIN}.

In the presence of an external magnetic field $B$, the energy density
functional is the sum of the standard Skyrme density functional
plus the interacting term between matter and  $B$,

\begin{equation}
{\mathcal E}=\sum_{\sigma}\, \frac{K_{\sigma}}{2 m^*_{\sigma}}+\frac{1}{16}\,a \,
(n^2 - W^2)- \kappa W B , \label{Enerdens}
\end{equation}
where
\begin{equation}
K_{\sigma}=\frac{1}{\Omega}\sum_{\vec k}k^2 f_{\sigma}({\vec k},
T) \label{Kindens}
\end{equation}
is the kinetic density, and we have introduced the effective nucleon
mass $m^*_{\sigma}$ for a definite spin polarization state, defined
as
\begin{equation}
\frac{1}{m^*_{\sigma}}=\frac{1}{m}+\frac{1}{4}\, ( b_0\,n + s \, b_1 \,
W) \label{SkmEffMass}
\end{equation}
with as before $s=1(-1)$for $\sigma=\uparrow(\downarrow)$.

The single-particle spectrum is,

\begin{equation}
\epsilon_{\sigma}({\vec k})=\frac{k^2}{2 m^*_{\sigma}}+\frac{1}{8}
\, v_{\sigma} \mp \kappa B \label{SPXpectrum}
\end{equation}
which is obtained in a self--consistent way through the functional
derivative $\epsilon_{\sigma}({\vec k})=\delta {\mathcal E}/\delta f_{\sigma}({\vec k},T)$.
In Eq.~(\ref{SPXpectrum}) we have used,

\begin{equation}
v_{\sigma}=a\, (n \mp W) + \sum_{\sigma'} ( b_0 +(2\delta_{\sigma\sigma'}-1) b_1)
K_{\sigma'}+\frac{\alpha}{3}  t_3(1-x_3)(n^2-W^2)n^{\alpha-1},
\label{SPPot}
\end{equation}
where the last term corresponds to the rearrangement contribution.

 The parameters $a,\, b_0$ and $b_1$ can be written in terms of the
standard parameters of the Skyrme model,
\begin{eqnarray}
a & = & 4 t_0 (1 - x_0)+ \frac{2}{3}t_3 (1 - x_3) n^\alpha \nonumber \\
b_0& = & t_1(1 - x_1)+ 3 t_2 (1 + x_2)\nonumber\\  b_1 & = & t_2
(1 + x_2) - t_1(1 - x_1). \nonumber
\end{eqnarray}
 we have
adopted $M=\kappa W$ for the magnetic moment of the system. For
given values of $n$, $T$ and $B$ we solve in a self-consistent way
the set of Eqs.~(\ref{Enerdens})-(\ref{SPPot}), obtaining the spin
polarization $W$ and the chemical potential $\mu$
that reproduces the density. The
physical state corresponds to that configuration (among CPS-U, 
CPS-D and PPS) which gives the lowest
value of ${\mathcal F}$.

We want to point out that here the physical state corresponds to a
minimum of ${\cal F}$, in contrast with previous investigations of
two of the authors \cite{UNLP}, where a transformed
thermodynamical potential was used.

%%%%%%%%%%%%%%%%%%%%%%%%%%%%%%%%%%%%%%%%%%%%%%%%%%%%
\section{Results and discussion}
\label{Results}

Before we present our results we would like to make a comment on the
validity and limitations of the models considered. Generally speaking, the validity of most
of the nuclear models is questionable in the limit of high densities and high isospin
asymmetries where the description of nuclear matter requires the inclusion of additional
degrees of freedom  and phenomena such as {\it e.g.,} hyperons, meson condensates or the chiral
and quark-gluon plasma phase transitions. In addition, the reader should also note that
in the case of non-relativistic models, as {\it e.g.,} BHF and Skyrme, causality is not always
guaranteed at high densities. To avoid such problems and to highlight the aim of this
work, we have restricted our calculations to densities below $2.5\,n_0$. The density and
temperature domain chosen ensures an {\it a priori} reasonable agreement
among the different theoretical descriptions. Taking into account the smallness of the neutron
magnetic moment, we also expect that even the largest magnetic field considered
in this work, $B=10^{19}$ G, will not modify essentially the dynamical
regime of the nuclear interactions.  We finish this comment by mentioning that, although,
the SLy4 parametrization considered here shows an anomalous spontaneous magnetization
at $n \simeq 4 \, n_0$ \cite{ISAEV2} we do not expect it to have any influence in the subsaturation
density region. Possible effects on the medium range densities will be timely commented.

In the following we discuss the results obtained for homogeneous
neutron matter under a strong magnetic field, with the models and approximations described in
the previous section. In all the figures, we show results corresponding to the physical
state, {\it i.e.,} that within the possible configurations CPS-U, CPS-D
and PPS which gives a minimum value of the free-energy density
${\mathcal F}$. As the density, temperature, and field intensity
changes, the system can pass from one global configuration to
another. For example, for fixed temperature and intensity $B$ the
system can pass from CPS-D to PPS as the density increases. In a
similar way, for fixed density and temperature, an increase of the
magnetic intensity leads the system from a PPS to a PPS-D.
We define as threshold density $n_t$ (threshold field $B_t$)
the value of the density (field) where the minimum free energy
${\mathcal F}$ changes from one state of polarization to another
for fixed values of $B$ and $T$ ($n$ and $T$).

\begin{figure}%[b]
 \vspace{0.5cm}
 \includegraphics[width=0.75\linewidth,angle=0]{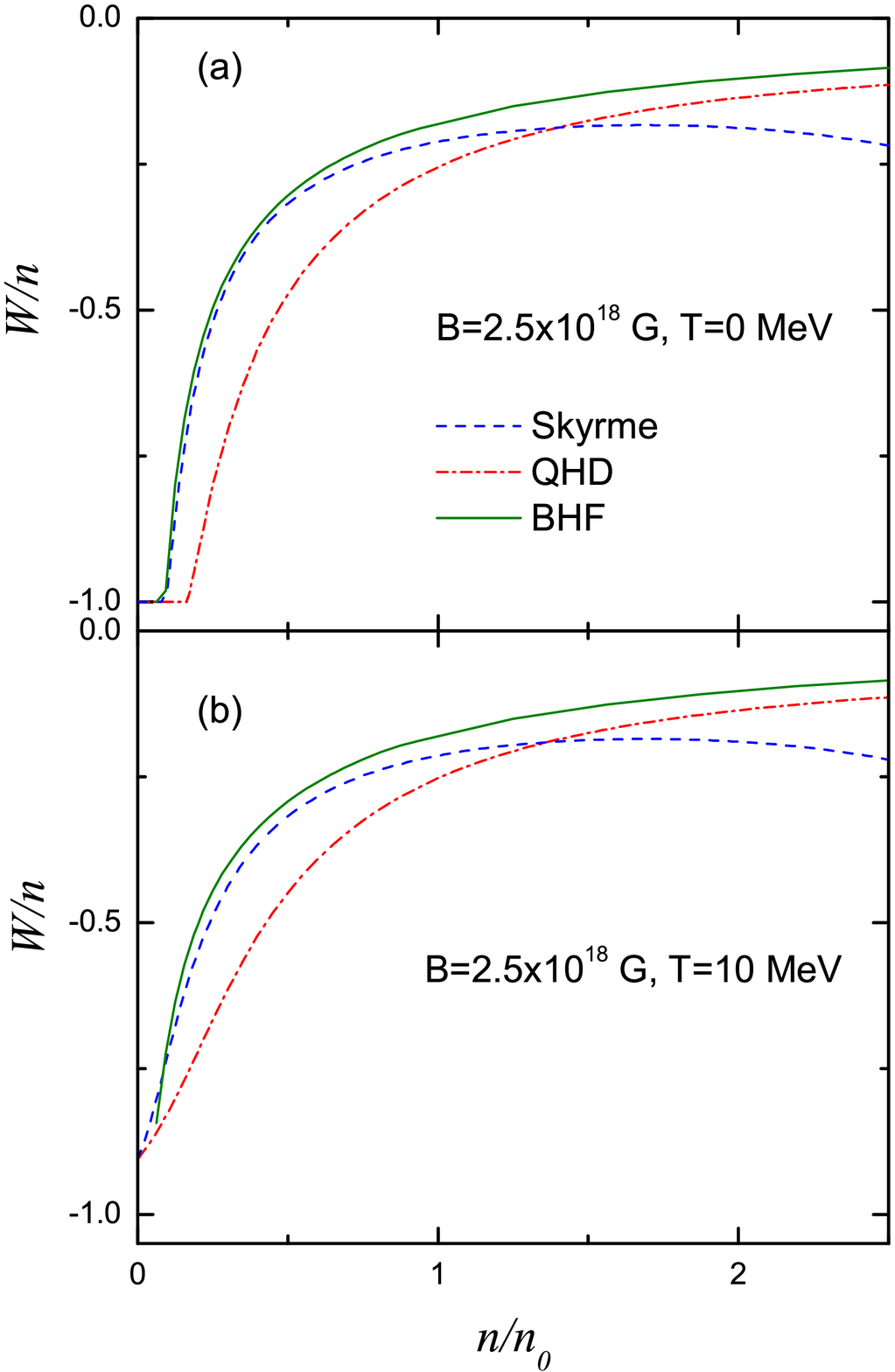}
\caption{(Color online) Spin asymmetry as a function of the density
at $B=2.5\times 10^{18}$ G and $T=0$ (panel a) and $T=10$ MeV
(panel b) for the three models considered.}
\label{fig1}
\end{figure}

\begin{figure}%[b]
 \vspace{0.5cm}
 \includegraphics[width=0.75\linewidth,angle=0]{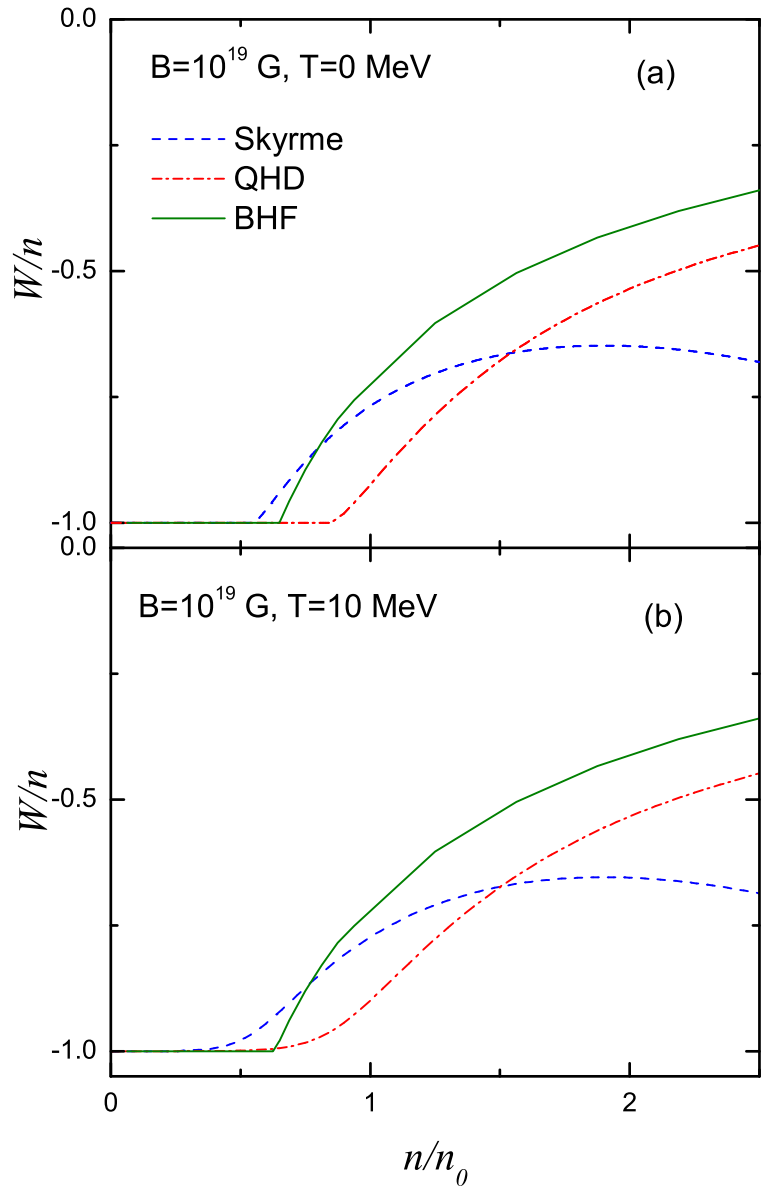}
\caption{(Color online) Same as Fig.\ \ref{fig1} for $B=10^{19}$ G.}
\label{fig2}
\end{figure}

We consider firstly the spin asymmetry density $W$, which gives us
information about the global state of polarization of the system.
In particular, Figs.\ \ref{fig1} and \ref{fig2} show the ratio
$W/n$ in terms of the total neutron density, different magnetic
field intensities and temperatures. At zero temperature (see
Figs.\ \ref{fig1}(a) and \ref{fig2}(a)) and for very low
densities, the system is completely polarized ($W/n=-1$) up to a
threshold density, $n_t$, where it changes to partially polarized,
with predominance of spin down states ($-1 < W/n < 0$).  A
comparison of the $B=2.5 \times 10^{18}$ G (Fig. \ref{fig1}) and
$B= 10^{19}$ G (Fig. \ref{fig2}) cases, shows that the threshold
density increases with $B$. However its precise location depends
on the model used.
For instance for $B=10^{19}$ G
at $T=0$ (see Fig.\ \ref{fig2}a),
we obtain $n_t/n_0=0.55$ for Skyrme,  $n_t/n_0=0.65$ for BHF,
and $n_t/n_0=0.85$ for QHD. It must be noted that beyond the
threshold, both BHF and QHD predicts always a monotonous growth,
reaching asymptotically the nonpolarized state ($W/n=0$) at high
densities. On the contrary, for the Skyrme model, the system is
always in a partially polarized state.
This behavior is a consequence of the well known ferromagnetic
instability predicted by the Skyrme model at high densities.

\begin{figure}%[b]
 \vspace{0.5cm}
 \includegraphics[width=0.75\linewidth,angle=0]{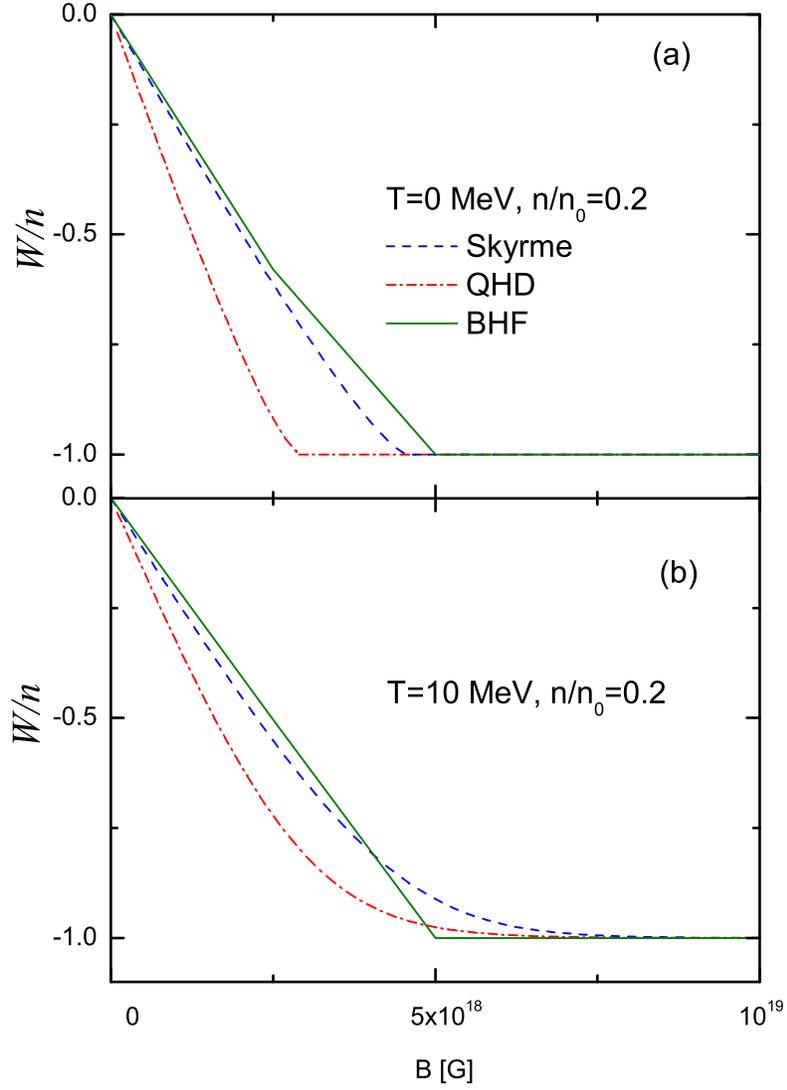}
\caption{(Color online) Spin asymmetry as a function of the magnetic
field intensity for $n/n_0=0.2$ and $T=0$ (panel a) and $T=10$
MeV (panel b) for the three models considered. The magnetic filed intensity is given
in units of $10^{18}$ G.}
\label{fig3}
\end{figure}

A similar description remains valid at higher
temperatures (see Figs.\ \ref{fig1}(b) and \ref{fig2}(b)), but
the passage from CPS to PPS becomes softer for
QHD and Skyrme. Hence, the
definition of a threshold density no longer makes sense for those
cases. Additional details can be seen in Fig.\ \ref{fig3}, where $W/n$ is
depicted as a function of $B$ for a fixed density $n/n_0=0.2$ and two
temperatures. Clearly, for $B=0$, there is no spin asymmetry
($W/n=0$) and the rate at which it changes from this value to a
completely polarized configuration ($W/n=-1$), is more pronounced
for QHD than for Skyrme, and for Skyrme than for BHF. At $T=0$ a
quick change of slope is detected at the transition point. The
BHF result keeps this feature still at $T=10$ MeV. At the same
temperature, the change from PPS to CPS-D, becomes a soft passage
for both QHD and Skyrme. From Figs.\ \ref{fig1}-\ref{fig3}, we see that
the temperature-dependence of the spin asymmetry is weaker for
BHF than for the other two models.

\begin{figure}%[b]
 \vspace{0.5cm}
 \includegraphics[width=0.75\linewidth,angle=0]{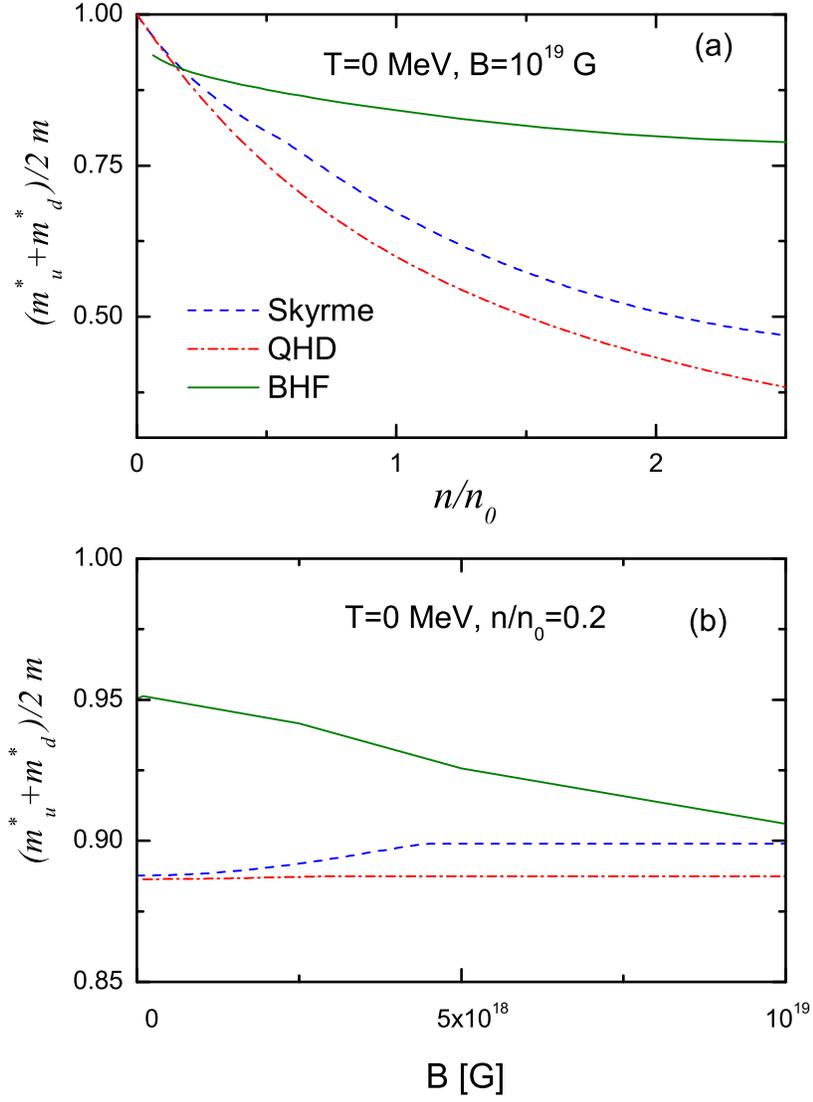}
\caption{(Color online) Panel a: average effective mass as a
function of the density at $T=0$ MeV for $B=10^{19}$G. Panel b:
average effective mass as a function of the magnetic field intensity
for $n/n_0=0.2$ and $T=0$. The magnetic filed intensity is given
in units of $10^{18}$ G.}
\label{fig4}
\end{figure}

The effects of an external magnetic field on the single-particle
properties can have significative consequences, for instance in the
transport properties in a dense nuclear medium. We examine in the
following the neutron effective mass, which is representative of the
single neutron properties.  In order to compare fairly the different
predictions, we define a spin averaged effective mass
$m^*=(m^*_\uparrow+m^*_\downarrow)/2m$ for BHF and Skyrme. It must
be taken into account that within the QHD model, $m^*$ is a scalar
which does not have an explicit dependence on the spin state. Let us
also recall that the effective mass has a momentum--dependence in
the BHF model and for purposes of comparison we fix $m_{\sigma}^*
\cong m_{\sigma}^*(k_{F_\sigma})$, where $k_{F_\sigma}$, is the
Fermi momentum of neutrons with spin projection  $\sigma$. This
comparison is presented in Fig.\ \ref{fig4}, where  $m^*$ is shown
as a function of the density at $T=0$ and $B=10^{19}$ G in Fig.\
\ref{fig4}(a), and as a function of the magnetic field intensity at
$T=0$ and $n/n_0=0.2$ in Fig.\ \ref{fig4}(b). In Fig.\ \ref{fig4}(a),
we found a monotonous
decrease over the range of densities studied here. For the QHD and
Skyrme models the spin average effective mass decreases
approximately  to one half of its vacuum value for $n/n_0=1.5$,
whereas in the BHF case  it exhibits a $\sim 20-22 \%$ decrease at
most.  As seen in the lower panel, the effect of the magnetic field
on $m^*$ is small for all the models at $n/n_0=0.2$. In particular,
for the QHD model $m^*$ is almost constant with $B$, whereas it
increases by about $\sim 1-2 \%$ for the Skyrme one, and decreases
by about $\sim 5 \%$ in the BHF case.

We have checked that for higher densities the magnetic effect on
$m^*$ becomes negligible for both the BHF and QHD models because
for these two models (see Figs.\ \ref{fig1} and \ref{fig2}) the
spin asymmetry $W/n$ goes to zero as density increases, the effect
of the magnetic field therefore being less and less important. On
the contrary, for the Skyrme model the effect $B$ on $m^*$ becomes
significative already for densities $n/n_0 > 0.3$, and it is
emphasized as the density grows. This is again a consequence of
the ferromagnetic instability predicted by the Skyrme model at
high densities mentioned before.

\begin{figure}%[b]
 \vspace{0.5cm}
 \includegraphics[width=0.75\linewidth,angle=0]{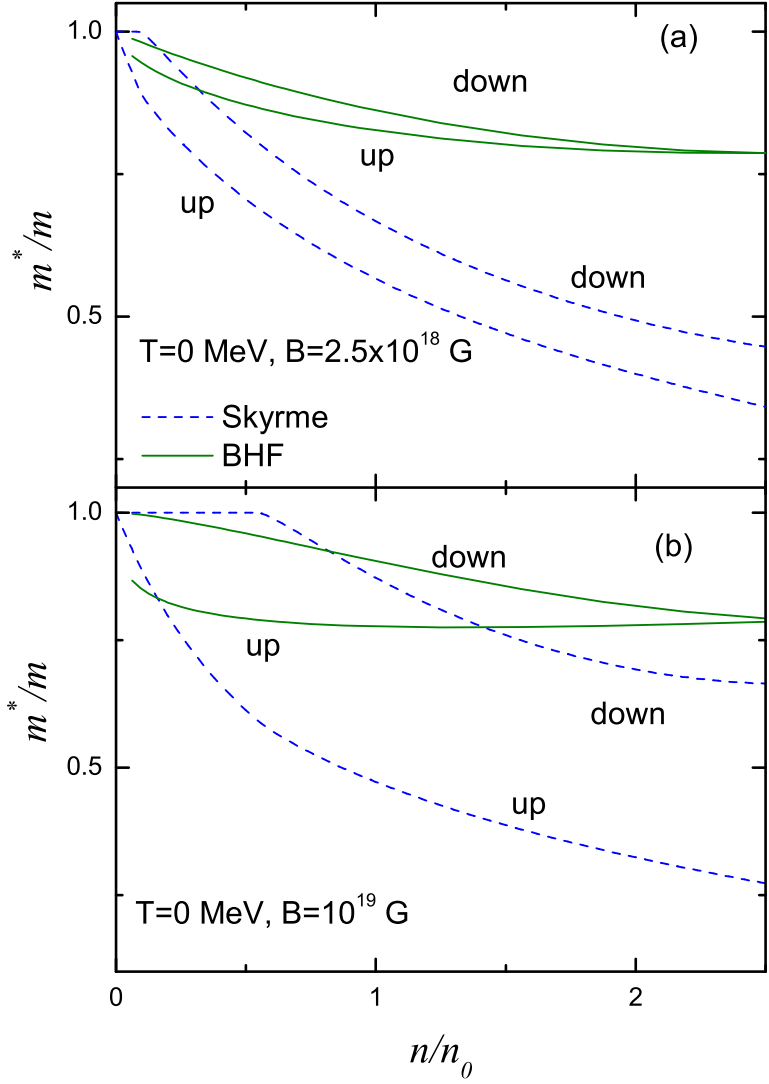}
\caption{(Color online) Neutron up and neutron down effective masses
as a function of the density for $T=0$ MeV and $B=2.5\times 10^{18}$
G (panel a) and $B=10^{19}$ G (panel b). Results are shown
only for the BHF and Skyrme models.}
\label{fig5}
\end{figure}

We analyze now the effective mass corresponding to different spin
polarization states within the non-relativistic potentials. The
dependence on the density at $T=0$, depicted in Fig.\ \ref{fig5}
shows some interesting features. In both cases $m_\downarrow^*$ is
larger than $m_\uparrow^*$, and the splitting, for a fixed density,
increases with $B$. For the BHF model, however, this difference
decreases as the density increases, and the two masses cross at
$n/n_0 \sim 2.5$. The reason is that in the BHF case, when density
increases, the effect of the magnetic field becomes less important
and is completely negligible when the system reaches the
nonpolarized state at high densities. On the other hand, the Skyrme
model shows a perceptible difference, even for extreme densities,
due to the ferromagnetic instability predicted by this model.
Furthermore, for densities $n<n_t$, $m_\downarrow^*$ saturates at
its vacuum value for Skyrme. This is a consequence of the particular
SLy4 parametrization used, for which is $b_{0}+b_{1}=0$ for neutrons with spin
down in the CPS-D state (see Eq.\ (\ref{SkmEffMass})).

\begin{figure}%[b]
 \vspace{0.5cm}
 \includegraphics[width=0.75\linewidth,angle=0]{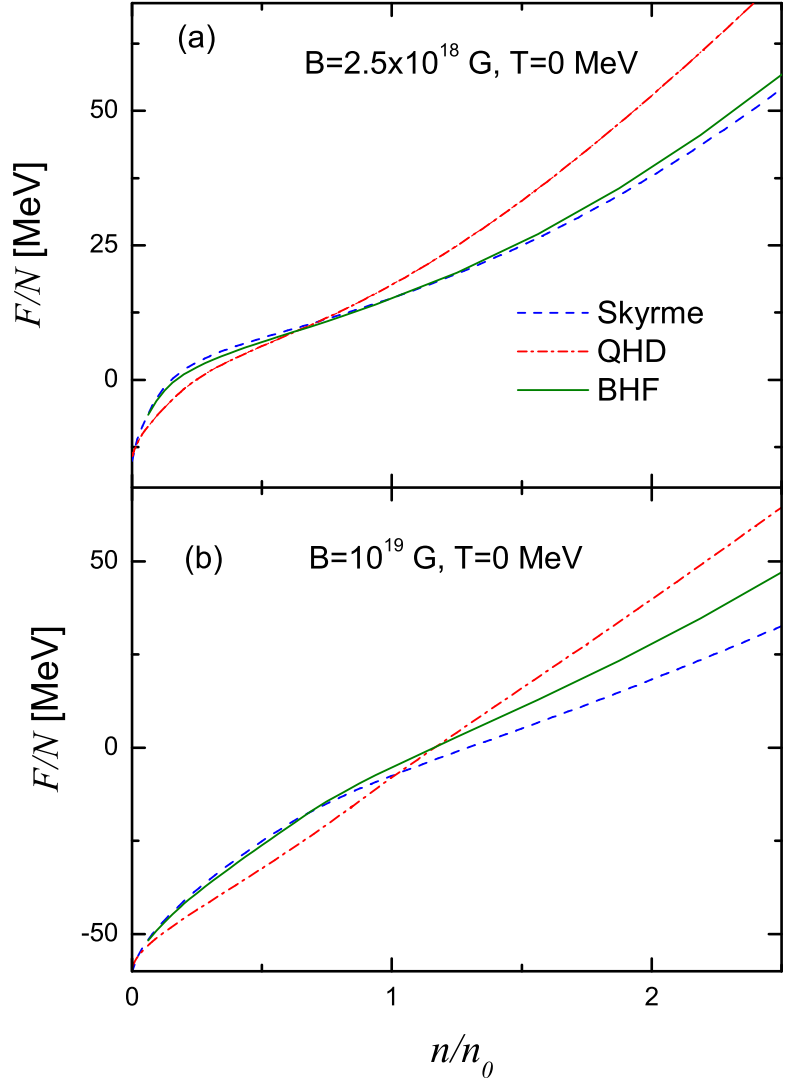}
\caption{(Color online) Free-energy per particle as a function of the density for $T=0$ MeV
and $B=2.5\times 10^{18}$ G (panel a) and $B=10^{19}$ G (panel b). }
\label{fig6}
\end{figure}

In the following we discuss some bulk thermodynamical properties.
The first one, is the free energy per particle, $F/N={\mathcal
F}/n$, shown in Fig.\ \ref{fig6} as a function of the density for
$T=0$ and for two magnetic field intensities. It must be mentioned
that, for the sake of comparison, the rest mass contribution was
subtracted in the QHD results.

It is a well known fact that neutron matter is not bound by the
action of nuclear forces. However, as seen in the figure, the
presence of a magnetic field of $\sim 2.5\times 10^{18}$ G leads to
a bound state, already at low densities. The binding increases when
the strength of the field grows, and for fields  $B \sim 10^{19}$
G neutron matter is bound up to saturation density. At relatively
low densities the kinetic energy and the repulsion between neutrons
are reduced, the effect of the magnetic field becoming the dominant
one. For medium and high densities the repulsive character of the
neutron-neutron interaction and the kinetic energy dominate over the
magnetic field and the system becomes unbound. We note that there is
good agreement between BHF and Skyrme for densities up to $n_{0}$.

\begin{figure}%[b]
 \vspace{0.5cm}
 \includegraphics[width=0.75\linewidth,angle=0]{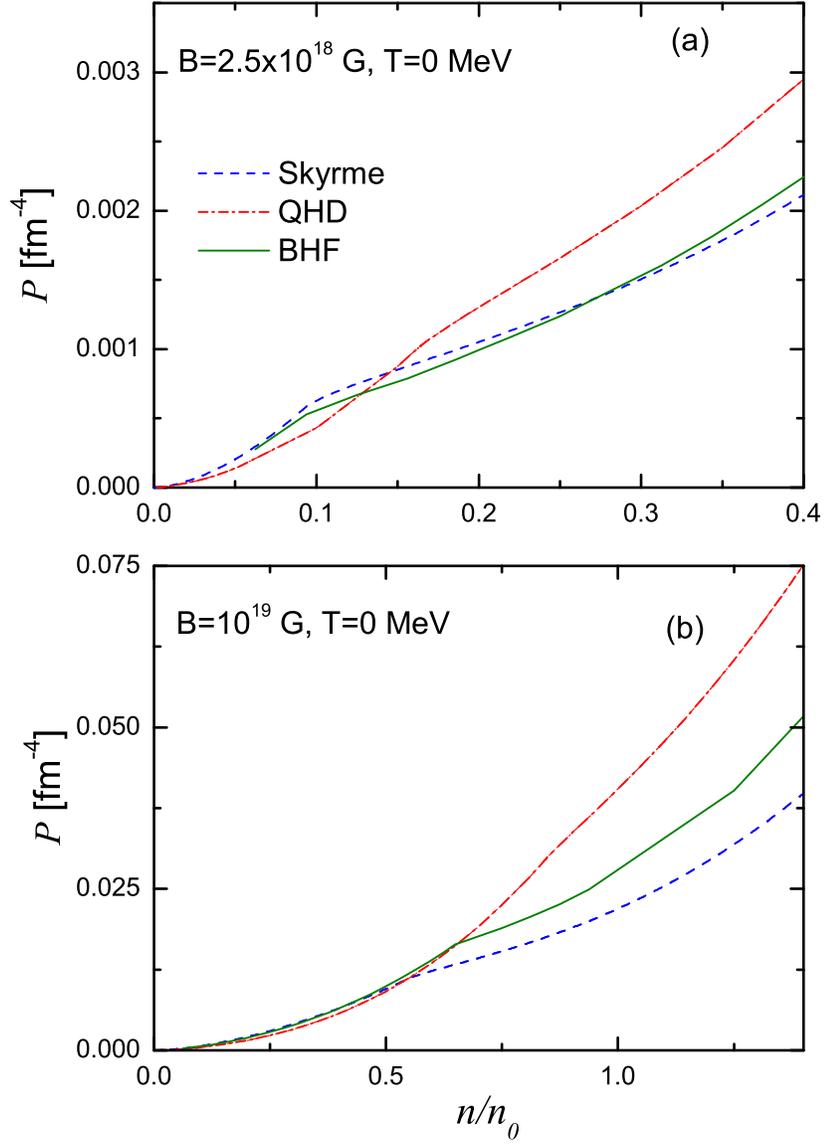}
\caption{(Color online) Pressure as a function of the density for $T=0$ MeV
and $B=2.5\times 10^{18}$ G (panel a) and $B=10^{19}$ G (panel b). }
\label{fig8}
\end{figure}

To carry on with the study of some bulk thermodynamical properties,
we focus now on the pressure. It is shown  as a function of the
density for T=0 in Fig.\ \ref{fig8}, where we have selected a
range of densities below the saturation value $n_0$ and two magnetic
field intensities. As is required by stability conditions the curves
show a monotonous increasing behavior. A careful inspection for all
the models shows a slight change of slope at the threshold
densities $n_t$, where the system changes its polarization state
from CPS-D to PPS. We have checked that the temperature variation
within the range covered in this work has no significative effects
on the pressure for any of the models.

Up to this moment we have obtained compatible descriptions of the
equation of state, without discontinuities and with some differences
in the values of the density where the system changes from CPS-D to
PPS. Hence, it is interesting to analyze some of the first
derivatives of the thermodynamical potentials. We choose as
significative examples the isothermal compressibility $K$, and the
magnetic susceptibility $\chi$, as stated in Sec. II.

\begin{figure}%[b]
 \vspace{0.5cm}
 \includegraphics[width=0.75\linewidth,angle=0]{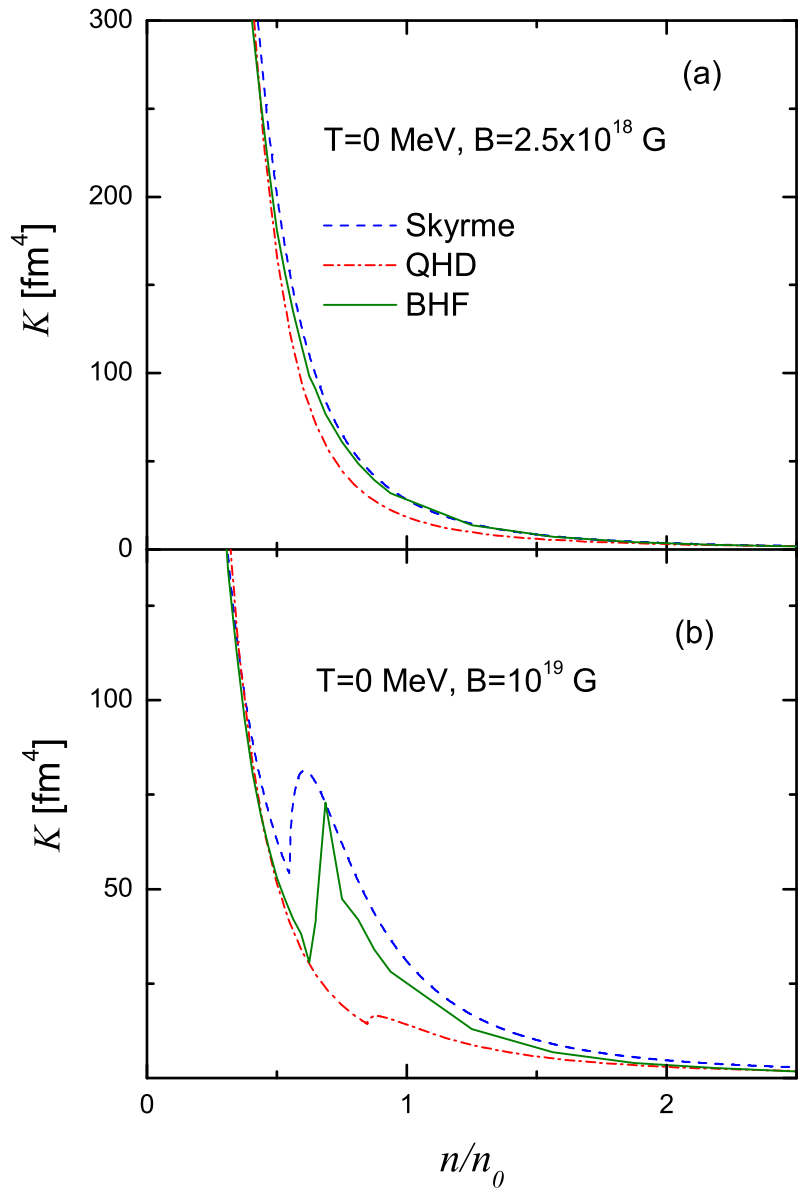}
\caption{(Color online) Isothermal compressibility as a function of
the density for $T=0$ MeV and $B=2.5\times 10^{18}$ G (panel a)
and $B=10^{19}$ G (panel b).}
\label{fig9}
\end{figure}

\begin{figure}%[b]
 \vspace{0.5cm}
 \includegraphics[width=0.75\linewidth,angle=0]{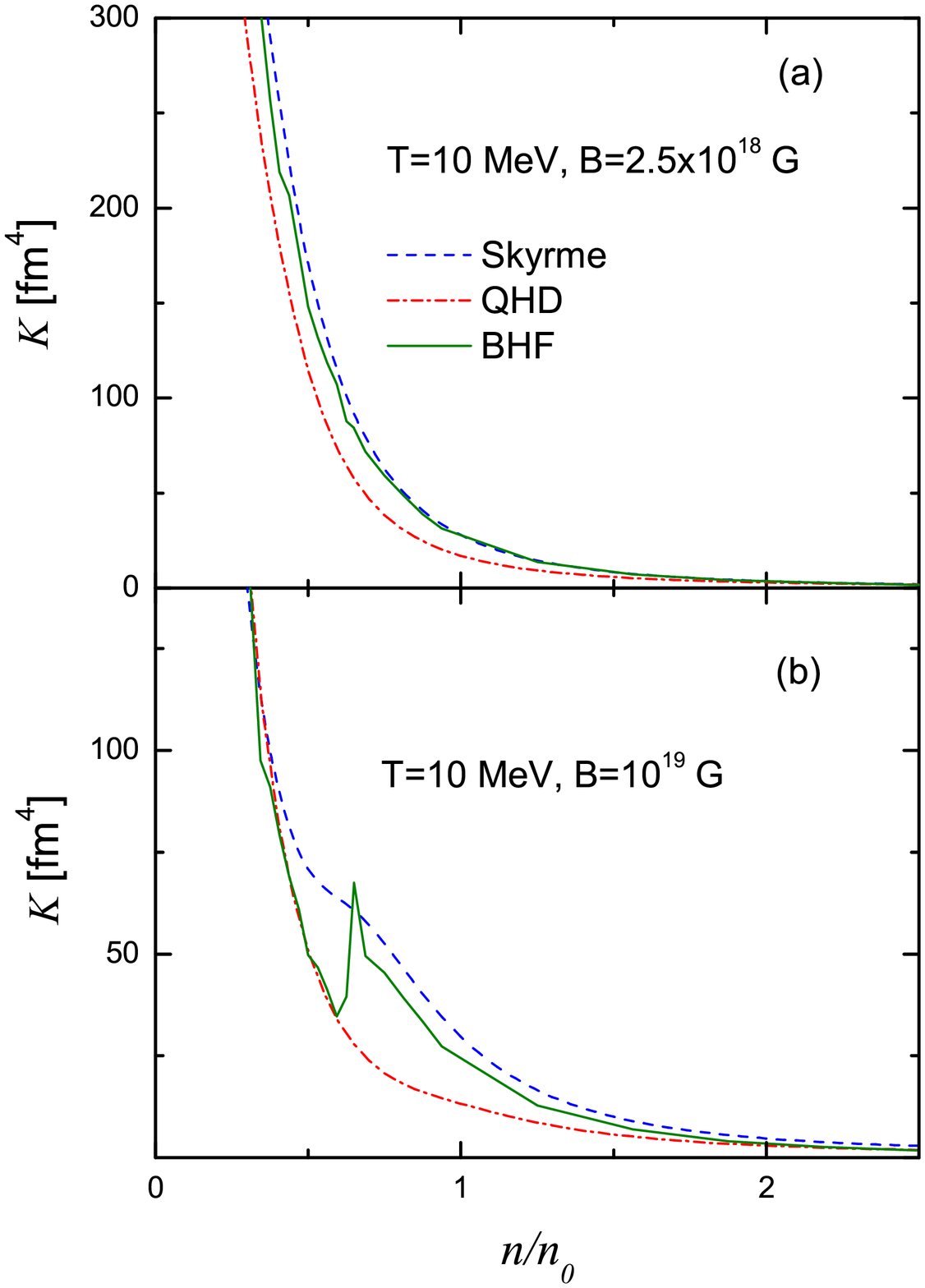}
\caption{(Color online) Same as Fig.\ \ref{fig9} for $T=10$ MeV. }
\label{fig10}
\end{figure}

In the first place, we show in Fig.\ \ref{fig9} the isothermal
compressibility $K$ as a function of the density at zero
temperature for $B=2.5\times 10^{18}$ G in Fig.\ \ref{fig9}(a) and
$B=10^{19}$ G in Fig.\ \ref{fig9}(b). For these magnetic
intensities the compressibility falls from relatively high values
at very low densities, decreasing monotonously with density until
a peak appears at the threshold density $n_t$. The
origin of this peak is due simply to the change of the slope of
the pressure at the threshold density $n_t$ (see Fig.\
\ref{fig8}). Beyond this point, the compressibility behaves in a
very similar way for all the models. From the asymptotic behavior
exhibited, it can be said that under the hypotheses assumed,
neutron matter can be considered incompressible for $n/n_0 > 2$.
We note also that for $B=2.5\times 10^{18}$ G the same kind of
peaks are present at very low densities but they are not visible
on this figure. In Fig.\ \ref{fig10} it is shown that thermal
effects smear out the peaks within the Skyme and QHD descriptions.
Note that the BHF result is almost insensitive
to thermal effects, showing a peak similar to that of the zero temperature
case.

\begin{figure}%[b]
 \vspace{0.5cm}
 \includegraphics[width=0.75\linewidth,angle=0]{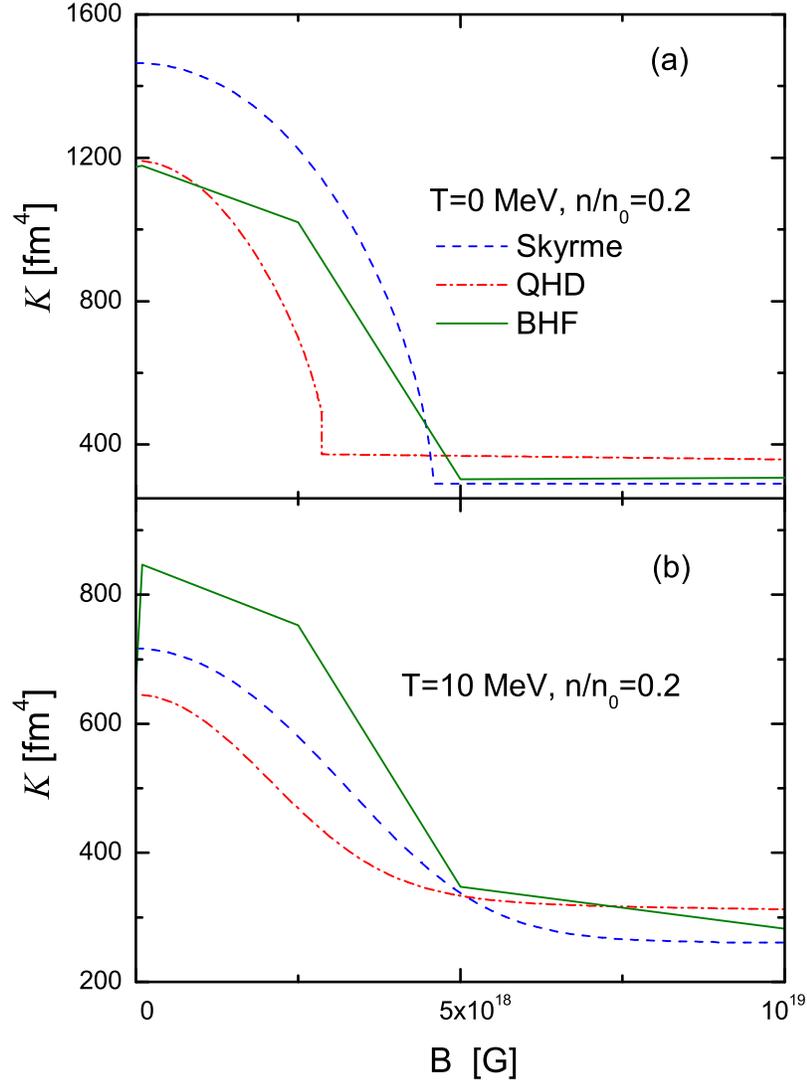}
\caption{(Color online) Isothermal compressibility as a function of
the magnetic field intensity for $n/n_0=0.2$ and $T=0$ MeV (panel a)
and $T=10$ MeV (panel b). The magnetic filed intensity is given
in units of $10^{18}$ G.}
\label{fig11}
\end{figure}

The dependence of $K$ on the magnetic field intensity at a fixed
density $n/n_0=0.2$ is exhibited in Fig.\ \ref{fig11}.  For $T=0$
(Fig.\ \ref{fig11}(a)) there are two different regimes,  in the
low field region the compressibility resembles an inverted
parabola. For stronger fields, $K$  reaches a plateau with an
almost constant value $K \sim 400$ fm$^4$. The
change of regime takes place at a threshold field intensity $B_t$, with
an abrupt change of slope.  The value of $B_t$ depends on the
model, the lower one, $B_t = 2.86 \times 10^{18}$ G, corresponds to
QHD whereas those of Skyrme and BHF are $B_t = 4.60 \times 10^{18}$ G
and $B_t = 5 \times 10^{18}$ G, respectively.
The plateau can be easily
understood by taking into account that for values of $B > B_t$
the system is completely polarized, {\it i.e.,} $W = -n$, and a further
increase of $B$ has no effect on it. Consequently, the value of $K$ remains
equal to that at $B=B_t$.
The increment of temperature
(see Fig.\ \ref{fig11}(b)) seems to erase this abrupt change of
slope for the Skyrme and QHD models, whereas the BHF case keeps
the angular points still for $T=10$ MeV. In addition, the
asymptotic values are smaller than the ones for $T=0$.

Note that the isothermal incompressibility ($K^{-1}$) was studied in Ref.
\cite{ANG3}, at $T=0$ and relatively low magnetic intensities. Using
the SLy7 parametrization of the Skyrme model, a monotonous behavior
was found for the low to medium densities regime.

\begin{figure}%[b]
 \vspace{0.5cm}
 \includegraphics[width=0.75\linewidth,angle=0]{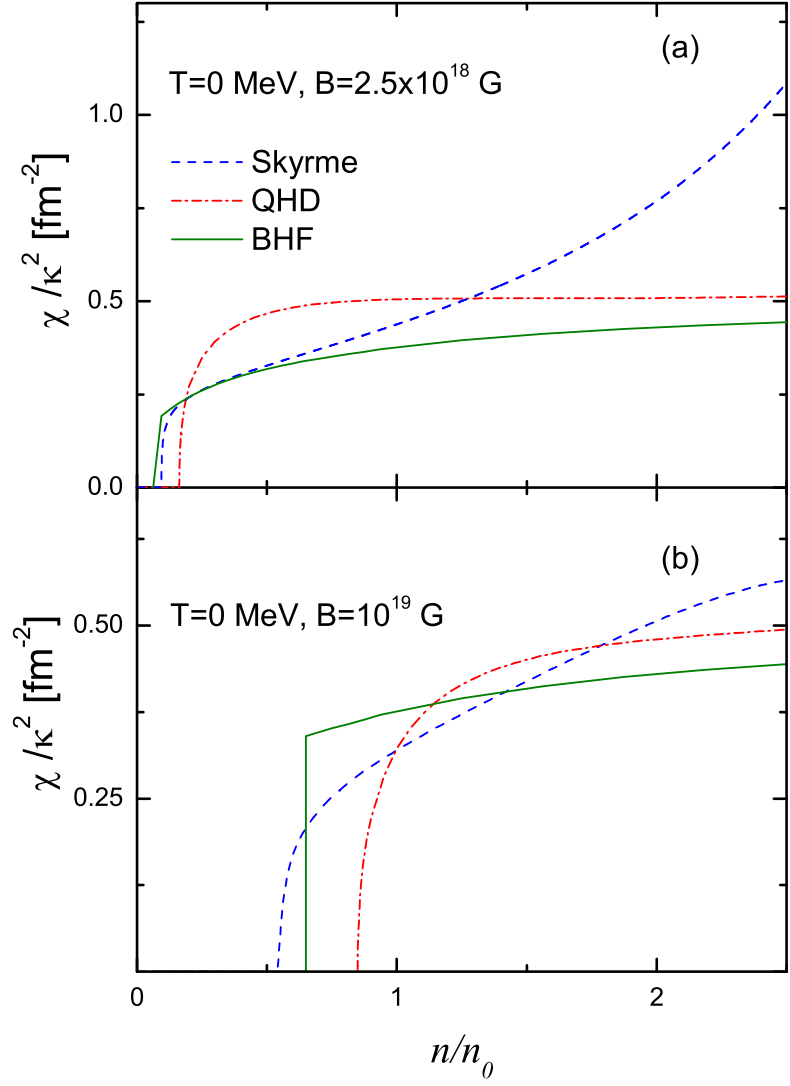}
\caption{(Color online) Magnetic susceptibility over $\kappa^2$ as a function of the
density for $T=0$ MeV and $B=2.5\times 10^{18}$ G (panel a) and
$B=10^{19}$ G (panel b).} \label{fig12}
\end{figure}

\begin{figure}%[b]
 \vspace{0.5cm}
 \includegraphics[width=0.75\linewidth,angle=0]{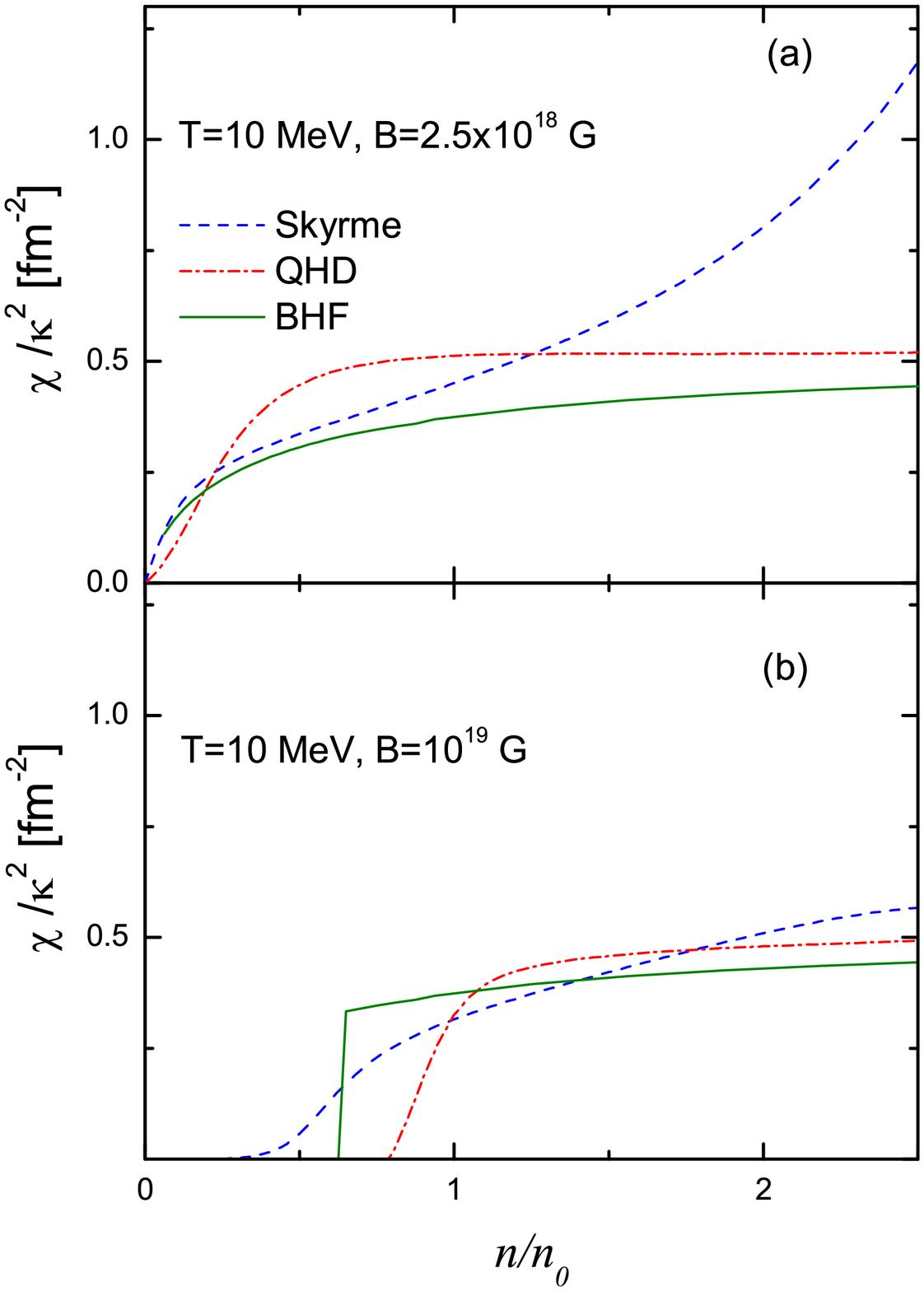}
\caption{(Color online) Same as Fig.\ \ref{fig12} for $T=10$ MeV.}
\label{fig13}
\end{figure}

Finally, we analyze the magnetic susceptibility, which is very
weak for neutron matter. However, as shown in the following,
it provides valuable information about the character of the change
of spin polarization. In Fig.\ \ref{fig12} the susceptibility is
shown as a function of the density for $T=0$ and two values of the
magnetic field. For densities smaller than the threshold density
$n_t$ and zero temperature the magnetization of the system is
saturated. Therefore, a further increase of the field intensity
does not change the magnetization.  Consequently, we have $\chi
=0$ for $n<n_t$. A sharp increase is detected for densities
slightly above $n_t$. Beyond that point, $\chi$ shows only
moderate variations in the QHD and BHF cases, whereas it grows
with an almost constant rate in the Skyrme one. In Fig.\ \ref{fig13},
we see that thermal effects,
as pointed out in similar circumstances discussed in this section,
smears out the abrupt changes in the QHD and Skyrme cases while it
seems to have a small effect in the BHF one.

\begin{figure}%[b]
 \vspace{0.5cm}
 \includegraphics[width=0.75\linewidth,angle=0]{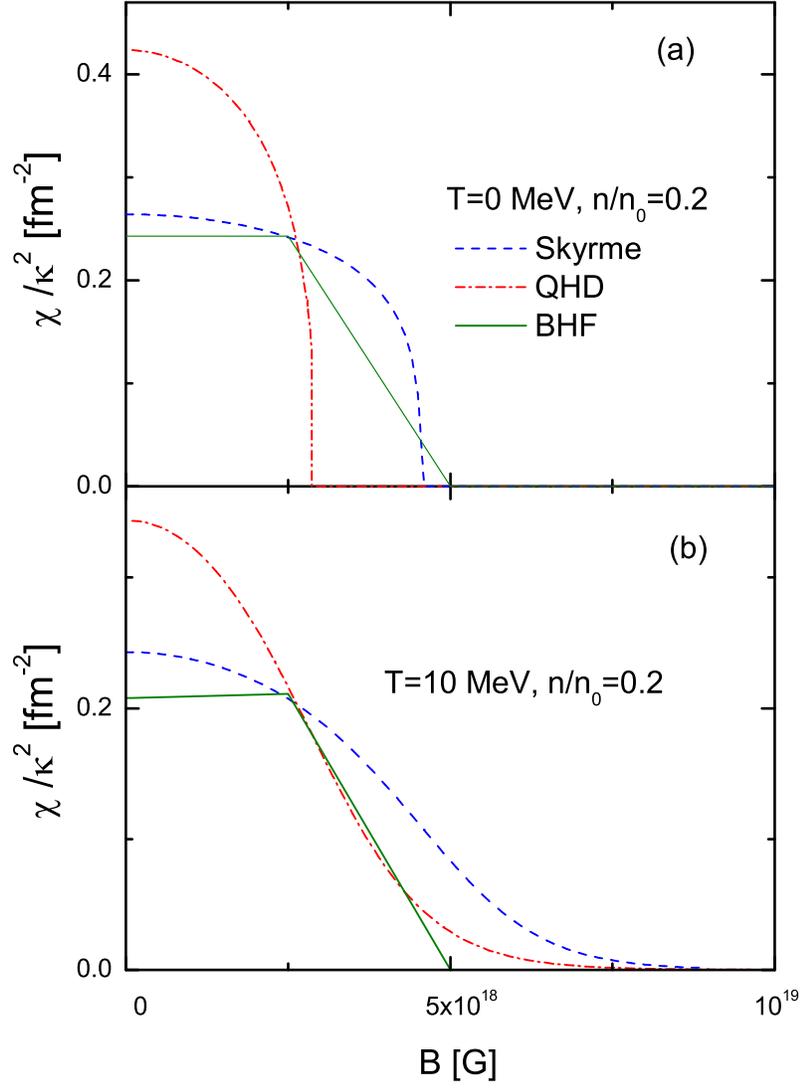}
\caption{(Color online)  Magnetic susceptibility over $\kappa^2$
as a function of the magnetic field intensity for $n/n_0=0.2$
and $T=0$ MeV (panel a) and $T=10$ MeV (panel b).
The magnetic filed intensity is given
in units of $10^{18}$ G.}
\label{fig14}
\end{figure}

The dependence of $\chi$ on the magnetic intensity is given in
Fig.\ \ref{fig14}. The density is fixed at $n/n_0=0.2$ and
temperature at $T=0$ (Fig.\ \ref{fig14}(a)) and $T=10$ MeV (Fig.\
\ref{fig14}(b)). A description similar to that given for Fig.\
\ref{fig11} holds here. As in that case, the same
threshold value $B_t$ separates, in each model, the high
intensity field regime, where $\chi/ \kappa^2 \simeq 0$, from the
monotonous decreasing trend found for the low field domain.
Since the susceptibility measures the rate of
change of the magnetization with the applied field, it is clear
that as the system saturates spin in the CPS-D state, the
susceptibility goes to zero. This fact explains, as in the case
of $K$, the plateau exhibited by $\chi$ for $B>B_t$.

\section{Conclusions}
\label{Summary}

In the present work we have analyzed the behavior of neutron
matter in the presence of an external magnetic field for a wide
range of densities, two temperatures and several magnetic field
intensities. Magnetic effects are small due to the smallness of
the intrinsic magnetic moment of the neutron. However, we have
found that there are some observables that give a clear signal of
a change in the physical configuration of the system in the low
density-low temperature regime. In order to give a discussion as
general as possible, we have used different models of the nuclear
interaction. All of them have been successfully used in different
fields of the nuclear physics, although they have very different
theoretical foundations. They are: {\it i}) the
Brueckner--Hartree--Fock (BHF) approach using the Argonne V18
nucleon-nucleon potential supplemented with the Urbana IX
three-nucleon force, {\it ii}) the covariant formulation known as
Quantum Hadro-dynamics (QHD) in its FSU-Gold version within a mean
field approach,  and {\it iii}) the SLy4 parametrization of the
non-relativistic Skyrme effective potential in a Hartree-Fock
scheme.

The spin asymmetry $W$ is a key feature to understand the behavior
of the system. The results for $W$ obtained by the different
models are in qualitative agreement. Within the range of magnetic field
intensities considered here, the system is completely polarized
for small densities up to a threshold density $n_t$, where it
changes into a partially polarized state. The value of $n_t$ increases with
the magnetic field intensity. There are some details, such as the
location of $n_t$, which differ from one model to the other.
However, they can be understood in terms of the in-medium nuclear
interaction. Thermal effects tend to soften the passage from
completely to partially polarized and reduce the degree of
polarization.

We have studied the effective mass due to its importance in the
single-particle properties.  We have found that it is a monotonous
decreasing function of the density for the three models. We have
seen that the effect of the magnetic field on $m^*$ is in general
small for all the models at low densities, becoming completely
negligible at high densities for the BHF and QHD models whereas, on
the contrary, for the Skyrme force it becomes more important as
density grows. This is a consequence of the well known
ferromagnetic instability predicted by these forces.

With regard to the equation of state, there are not significative
differences among the various predictions and only weak clues
about the change of polarization. The second derivatives of the
thermodynamical potentials, such as the compressibility and the
magnetic susceptibility, give clear evidence of a change in the
system. At zero temperature they show an abrupt change of regime
that becomes diffuse as the temperature is increased in the QHD
and Skyrme cases. The isothermal compressibility, for example, has
a non-monotonous behavior around the threshold density $n_t$. This
feature can have significative consequences as, for instance, in
the propagation of density waves through the crust of neutron
stars.

In conclusion, we have found robust results supported by the three
models. The change in the global polarization of the system does
not produce discontinuities in the thermodynamical potentials. The
remarkable change of the slope found in the equation of state at
the threshold point resembles a second order phase transition.
However, a detailed examination of the relevant second order
derivatives of the thermodynamic potential does not show any
discontinuity. Hence, we conclude the system undergoes a
continuous passage or experiences a higher order phase transition.
The consequences of the non-monotonous behavior of the
compressibility near the transition point requires further
investigation.

To establish significative differences among the
three models within the subject under study, additional
information must be taken into account. We can mention here the
cooling rate of a neutron star, which strongly depends on the
magnetization state of matter. In~\cite{ANG2} it was shown that
there is a decrease of the neutrino opacity of magnetized matter
with respect to the non-magnetized case. Of course, a realistic
description of this issue requires some refinements, such as the
inclusion of protons, leptons, and exotic degrees of freedom such
as hyperons in $\beta$-equilibrium. This would be
the natural extension of the present work and will be considered in
a near future. However, we believe that a good understanding of
the simpler neutron matter case is the first step in such
direction.

\newpage
\section*{Acknowledgements}This work was partially supported by the
CONICET, Argentina, under contracts PIP 0032 and PIP
11220080100740, by the Agencia Nacional de Promociones Cientificas
y Tecnicas, Argentina, under contract PICT-2010-2688, by the
initiative QREN financed by the UE/FEDER through the Programme
COMPETE under the projects PTDC/FIS/113292/2009 and
CERN/FP/123608/2011, and by the COST action MP1304 ``NewCompStar:
Exploring fundamental physics with compact stars''.

%%%%%%%%%%%%%%%%%%%%%%%%%%%%%%%%%%%%%%%%%%%%%%%%%%%%%%%%%%%

%%%%%%%%%%%%%%%%%%%%%%%%%%%%%%%%%%%%%%%%%%%%%%%%%%%%%%%%%%%%%%%%%%%%%%%%%%%%%%
\newpage
%%%%%%%%%%%%%%%%%%%%%%%%%%%%%%%%%%%%%%%%%%%%%%%%%%%%%%%%%%%%%%%%%%%%%%%%%%%%%%

%%%%%%%%%%%%%%%%%%%%%%%%%%%%%%%%%%%%%%%%%%%%%%%%%%%%%%%%%%%%%%%%%%%%%%%%%%%%%%
%%%%%%%%%%%%%%%%%%%%%%%%%%%%%%%%%%%%%%%%%%%%%%%%%%%%%%%%%%%%%%%%%%%%%%%%%%%%%%

%%%%%%%%%%%%%%%%%%%%%%%%%%%%%%%%%%%%%%%%%%%%%%%%%%%%%%%%%%%%%%%%%%%%%%%%%%%%%%
%%%%%%%%%%%%%%%%%%%%%%%%%%%%%%%%%%%%%%%%%%%%%%%%%%%%%%%%%%%%%%%%%%%%%%%%%%%%%%

%%%%%%%%%%%%%%%%%%%%%%%%%%%%%%%%%%%%%%%%%%%%%%%%%%%%%%%%%%%%%%%%%%%%%%%%%%%%%%
%%%%%%%%%%%%%%%%%%%%%%%%%%%%%%%%%%%%%%%%%%%%%%%%%%%%%%%%%%%%%%%%%%%%%%%%%%%%%%

%%%%%%%%%%%%%%%%%%%%%%%%%%%%%%%%%%%%%%%%%%%%%%%%%%%%%%%%%%%%%%%%%%%%%%%%%%%%%%
%%%%%%%%%%%%%%%%%%%%%%%%%%%%%%%%%%%%%%%%%%%%%%%%%%%%%%%%%%%%%%%%%%%%%%%%%%%%%%

%%%%%%%%%%%%%%%%%%%%%%%%%%%%%%%%%%%%%%%%%%%%%%%%%%%%%%%%%%%%%%%%%%%%%%%%%%%%%%
%%%%%%%%%%%%%%%%%%%%%%%%%%%%%%%%%%%%%%%%%%%%%%%%%%%%%%%%%%%%%%%%%%%%%%%%%%%%%%

%%%%%%%%%%%%%%%%%%%%%%%%%%%%%%%%%%%%%%%%%%%%%%%%%%%%%%%%%%%%%%%%%%%%%%%%%%%%%%
%%%%%%%%%%%%%%%%%%%%%%%%%%%%%%%%%%%%%%%%%%%%%%%%%%%%%%%%%%%%%%%%%%%%%%%%%%%%%%

%%%%%%%%%%%%%%%%%%%%%%%%%%%%%%%%%%%%%%%%%%%%%%%%%%%%%%%%%%%%%%%%%%%%%%%%%%%%%%
%%%%%%%%%%%%%%%%%%%%%%%%%%%%%%%%%%%%%%%%%%%%%%%%%%%%%%%%%%%%%%%%%%%%%%%%%%%%%%

%%%%%%%%%%%%%%%%%%%%%%%%%%%%%%%%%%%%%%%%%%%%%%%%%%%%%%%%%%%%%%%%%%%%%%%%%%%%%%
%%%%%%%%%%%%%%%%%%%%%%%%%%%%%%%%%%%%%%%%%%%%%%%%%%%%%%%%%%%%%%%%%%%%%%%%%%%%%%

%%%%%%%%%%%%%%%%%%%%%%%%%%%%%%%%%%%%%%%%%%%%%%%%%%%%%%%%%%%%%%%%%%%%%%%%%%%%%%
%%%%%%%%%%%%%%%%%%%%%%%%%%%%%%%%%%%%%%%%%%%%%%%%%%%%%%%%%%%%%%%%%%%%%%%%%%%%%%

%%%%%%%%%%%%%%%%%%%%%%%%%%%%%%%%%%%%%%%%%%%%%%%%%%%%%%%%%%%%%%%%%%%%%%%%%%%%%%
%%%%%%%%%%%%%%%%%%%%%%%%%%%%%%%%%%%%%%%%%%%%%%%%%%%%%%%%%%%%%%%%%%%%%%%%%%%%%%

%%%%%%%%%%%%%%%%%%%%%%%%%%%%%%%%%%%%%%%%%%%%%%%%%%%%%%%%%%%%%%%%%%%%%%%%%%%%%%
%%%%%%%%%%%%%%%%%%%%%%%%%%%%%%%%%%%%%%%%%%%%%%%%%%%%%%%%%%%%%%%%%%%%%%%%%%%%%%

%%%%%%%%%%%%%%%%%%%%%%%%%%%%%%%%%%%%%%%%%%%%%%%%%%%%%%%%%%%%%%%%%%%%%%%%%%%%%%
%%%%%%%%%%%%%%%%%%%%%%%%%%%%%%%%%%%%%%%%%%%%%%%%%%%%%%%%%%%%%%%%%%%%%%%%%%%%%%

\end{document}